\documentstyle[epsfig,twocolumn]{article}
\begin{document}

\newcommand{\be}{\begin{equation}}
\newcommand{\ee}{\end{equation}}
\newcommand{\bea}{\begin{eqnarray}}
\newcommand{\eea}{\end{eqnarray}}
\newcommand{\nn}{\nonumber}

\newcommand{\sot}{SO(3, \mathbb{R})}

\setcounter{page}{1}

\title{$SU(2)$ Yang-Mills quantum mechanics of spatially constant fields}
\author{H.-P. Pavel \\
        Institut f\"ur Kernphysik, c/o Theorie,
        Schlossgartenstr. 9,
        D-64289 Darmstadt, Germany \\
        and \\
        Bogoliubov Laboratory of Theoretical Physics,
        Joint Institute for Nuclear Research, Dubna, Russia\\
        }
\date{January 30, 2007}
\maketitle
\begin{abstract}

As a first step towards a strong coupling expansion of Yang-Mills theory, the $SU(2)$
Yang-Mills quantum mechanics of spatially constant gauge fields
is investigated in the symmetric gauge, with the six physical fields represented in terms
of a positive definite symmetric \(3\times 3\)  matrix \( S \).
Representing the eigenvalues of $S$ in terms of elementary symmetric polynomials,
the eigenstates of the corresponding harmonic oscillator problem can be calculated
analytically  and used as orthonormal basis of trial states for a variational calculation
of the Yang-Mills quantum mechanics. In this way
high precision results are obtained in a very effective way for the lowest eigenstates
in the spin-0 sector as well as for higher spin.
Furthermore I find, that practically all excitation energy
of the eigenstates, independently of whether it is a vibrational or a rotational excitation,
leads to an increase of the expectation value of the largest eigenvalue $\langle\phi_3\rangle$,
whereas the expectation values of the other two eigenvalues,
$\langle\phi_1\rangle$ and $\langle\phi_2\rangle$, and
also the component $\langle B_3\rangle=g\langle\phi_1\phi_2\rangle$ of the magnetic field,
remain at their vacuum values.

\end{abstract}


\section{\large\bf Introduction}

The so-called Yang-Mills mechanics originates from Yang-Mills field theory
under the supposition  of the spatial homogeneity of the gauge  fields \cite{BasMatSav}.
For this case  the Lagrangian of pure $SU(2)$ Yang-Mills theory
reduces to
\footnote{ Everywhere in the paper we put the spatial volume $V= 1$.
 As result the coupling constant $g$ becomes dimensionful
 with $g^{2/3}$ having the dimension of energy. The volume dependence
 can be restored in the final results by replacing $g^2$ with $g^2/V$. }
\be \label{hl}
L={1\over 2}\left(\dot{A}_{ai}-g\epsilon_{abc}A_{b0} A_{ci}\right)^2
   -{1\over 2} B_{ai}^2(A) ~,
\ee
with  the magnetic field
$B_{ai}(A)= (1/2)g\epsilon_{abc}\epsilon_{ijk}A_{bj}A_{ck}$.
The local $SU(2)$ gauge invariance and the rotational invariance
of the original Yang-Mills action
reduces to the symmetry under the $SO(3)$ transformations local in time
\bea
 A^{\omega}_{a0}(t)\!\!\!&=&\!\!\!
O(\omega(t))_{ab}A_{b0}(t) -\frac{1}{2g}
\epsilon_{abc}\left(O(\omega(t))\dot O(\omega(t)) \right)_{bc}\,,\nn\\
 A^{\omega}_{ai}(t)\!\!\!&=&\!\!\!
O(\omega(t))_{ab}A_{bi}(t)~,
\label{tr}
\eea
and the global spatial rotations
$A^{\chi}_{ai}=A_{aj}R(\chi)_{ji}$.

In the constrained Hamiltonian formulation (see e.g.\cite{Christ and Lee})
the time dependence of the gauge transformations (\ref{tr}) is exploited
to put the Weyl gauge $A_{a0} = 0~$, a=1,2,3, and
the physical states $\Psi$ have to satisfy both the Schr\"odinger equation
and the three Gauss law constraints
\bea
H\Psi &=& {1\over 2}
\sum_{a,i} \left[\left(\frac{\partial}{\partial A_{ai}}\right)^2+B_{ai}^2(A)\right]\Psi=E\Psi~,\\
G_a\Psi &=& -i\epsilon_{abc}A_{bi}\frac{\partial}{\partial A_{ci}}\Psi=0~,\quad a=1,2,3~.\label{G_a}
\eea
The $G_a$ are the generators of the residual time independent gauge transformations,
satisfying $[G_a,H]=0$ and $[G_a,G_b]=i\epsilon_{abc}G_c$.
Furthermore $H$ commutes with the angular momentum operators
$J_i  =  -i\epsilon_{ijk}A_{aj}\partial/\partial A_{ak}~, i=1,2,3$.
The matrix element of an operator $O$ is given in the Cartesian form
\be
\langle \Psi'| O|\Psi\rangle\
\propto
\int dA\
\Psi'^*(A) O \Psi(A)~.
\ee
For carrying out quantum mechanical calculations it is desirable to
have a corresponding unconstrained Schr\"odinger equation and to find its eigenstates
in an effective way with high accuracy at least for the lowest states.
The basic ideas and first results of such an gauge fixed approach
to Yang-Mills quantum mechanics will be presented in the following.

\section{\large\bf Unconstrained Hamiltonian }

\subsection{\normalsize\bf The symmetric gauge}

The local symmetry transformation (\ref{tr}) of the gauge potentials
\( A_{ai} \) promts us with the set of coordinates in terms of which the
separation of  the  gauge degrees of freedom occurs.
As in \cite{KP} I use the polar decomposition for arbitrary
\(3\times 3\) quadratic matrices
\be
\label{eq:pcantr}
A_{ai} \left(q, S \right)
= O_{ak}\left( q \right) S_{ki}~,
\ee
with the orthogonal matrix \( O (q)  \),
parametrized by the three angles \(q_i\), and the positive definite,
symmetric \(3\times 3\)  matrix \( S \). The decomposition
(\ref{eq:pcantr}) is unique and corresponds to the symmetric gauge
$\chi_i(A)\!\!=\!\!\epsilon_{ijk}A_{jk}\!\!=\!\!0$.
The Jacobian  is
$|\partial(A_{ai})/\partial(q,S)|\
\propto \det\Omega(q)\prod_{i<j}\left(\phi_i+\phi_j\right)$,
where $\phi_1,\phi_2,\phi_3$ are the eigenvalues of $S$ and
$\Omega_{jm}(q) \equiv (1/2) \epsilon_{mkl}
\left[O^T (q)\partial O\left(q\right)/\partial q_j\right]_{kl}$.
The variables \( S \) and \(\partial/\partial S  \) make no contribution to the
Gauss law operators
$G_a = -iO_{as}(q) \Omega^{-1}_{\ sj}(q)\partial/\partial q_j$.
Hence, assuming the invertibility of  the matrix
$ \Omega$, the non-Abelian Gauss laws (\ref{G_a})  are
equivalent to  the set of Abelian  constraints
$\partial \Psi/\partial q_i  = 0~, i=1,2,3$.
and the physical  Hamiltonian of $SU(2)$ Yang-Mills quantum mechanics reads
\bea
\label{eq:uncYMP}
 H  &=& {1\over 2}
\sum_{m,n} \Bigg[
-\left(\frac{\partial}{\partial S_{mn}}\right)^2  +
{1\over 2}\gamma^{-2}_{mn} J_m J_n +B_{mn}^2(S)\nn\\
&&\quad -\left[\gamma^{-1}_{mn}(S)-\delta_{mn}\mbox{tr}(\gamma^{-1}(S))\right]
    \frac{\partial}{\partial S}_{mn}~,
\Bigg]
\eea
with
$\gamma_{ik}(S) : = S_{ik} -  \delta_{ik}\mbox{tr} S$
and the angular momenta
$J_i= -2i\epsilon_{ijk}S_{aj}\partial/\partial S_{ak}~, i=1,2,3$,
in terms of the physical variables (note the factor $2$).
The matrix element of a physical operator O is given by
\be
\langle \Psi'| O|\Psi\rangle\
\propto
\int dS
\Big[\prod_{i<j}\left(\phi_i+\phi_j\right)\Big]
\Psi'^*(S) O\Psi(S)~.
\ee

\subsection{\normalsize\bf Unconstrained Hamiltonian in terms of
rotational and scalar degrees of freedom}

In order to achieve a more transparent form for the reduced Yang-Mills
system (\ref{eq:uncYMP})
I shall limit myself in this work to the principle orbit configurations
\be
\label{range}
0<\phi_1<\phi_2<\phi_3<\infty~,
\ee
for the eigenvalues $\phi_1,\phi_2,\phi_3>0$ of the positive definite symmetric matrix $S$
(not considering singular orbits where two or more eigenvalues coincide)
and perform a principal-axes transformation
\be
\label{patransf}
S  =  R(\alpha,\beta,\gamma)\ \mbox{diag}\ ( \phi_1 , \phi_2 , \phi_3 ) \
R^{T}(\alpha,\beta,\gamma)~,
\ee
with the \( SO(3)\) matrix  \({R}\) parametrized by the three Euler angles.
The Jacobian of  (\ref{patransf}) is
$|\partial S/\partial(\alpha,\beta,\gamma,\phi)| \propto
\sin\beta \prod_{i<j}\left(\phi_i- \phi_j\right)$.
In terms of the principal-axes variables, the physical Hamiltonian reads
\bea
\label{Hpys}
H\!&=&\!{1\over 2}\!\sum_{\rm
cyclic}^3\!\! \Bigg[-{\partial^2\over\partial \phi_i^2} -{2\over \phi_i^2-
\phi_j^2}\!\left(\phi_i{\partial\over\partial
\phi_i}-\phi_j{\partial\over\partial \phi_j}\right)\!\!\nn\\
&&\quad\quad\quad\quad +\xi_i^2{\phi_j^2+\phi_k^2\over (\phi_j^2-\phi_k^2)^2}
+\!  g^2 \phi_j^2 \phi_k^2\Bigg]~.
\eea
All the rotational variables in the Hamiltonian (\ref{HHHq}),
which agrees with that of \cite{Simon},
are combined into the intrinsic angular momenta $\xi_i=R^T_{ij}J_j$
obeying  $[\xi_i,\xi_j]=-i\epsilon_{ijk}\xi_k$.
The matrix elements of an operator $O$ are given as
\be
\langle\Psi'|O|\Psi\rangle\! \propto\!\!
\int\!\! d\alpha \sin\beta d\beta d\gamma\!\!\!\!\!\!\!\!\!\!\!\!\!
\int\limits_{0<\phi_1<\phi_2<\phi_3}\!\!\!\!\!\!\!\!\!\!\!\!\! \Big[\!\!\prod^{\rm cyclic}\! d\phi_i
\! \left(\phi_j^2- \phi_k^2\right)\Big]
\Psi'^* O\Psi.
\label{measure}
\ee
The potential term in (\ref{H--0}),
has a flat valley of degenerate absolute minima at zero energy,
$\phi_1=\phi_2=0,\phi_3-{\rm arbitrary}$,
at the edge of (\ref{range}).
Close to the bottom of the  valley the
potential is that of a harmonic oscillator with a
width narrowing down for larger values of $\phi_3$.

\subsection{\normalsize\bf Symmetries }

As a relic of the rotational invariance of the initial gauge field theory
the Hamiltonian (\ref{HHHq}) possesses the symmetry
\be
\label{cHI}
[H,J_k]=0~,
\ee
with the angular momentum operators $J_i= R_{ij}\xi_j$  satisfying
$[J_i, J_j] = i \epsilon_{ijk} J_k$ and  $[J_i, \xi_j] = 0~$.
Hence the eigenstates of $H$ can be characterized by the quantum numbers $J$ and $M$.
Furthermore $H$ is invariant under arbitrary permutations $\sigma_{ij}$ of any two of the
three indices $1,2,3$,
time reflections T, and parity reflections $ P: \phi_i\rightarrow -\phi_i $,
\be
[H,\sigma_{ij}]=0~,\ \ \ \ \ \ [H,T]=0~,\ \ \ \ \ \ [H,P]=0~.
\ee

\subsection{\normalsize\bf Boundary conditions}

The requirement of Hermiticity of $H$
in the region bounded by the three boundary planes $\phi_1=0~,\phi_1=\phi_2~, \phi_2=\phi_3~$
and at positive infinity, leads to the conditions at $\phi_1$
\be
\label{bc1}
\partial_1\Psi\Big|_{\phi_1=0}= 0
\quad\quad  \vee  \quad\quad
\Psi\Big|_{\phi_1=0}= 0  ~,
\ee
corresponding to even or odd parity states respectively.
I shall consider in the work only the first alternative allowing
only even parity states in all spin sectors.
Furthermore, due to the vanishing of the measure (\ref{measure}) on the
boundaries $\phi_1=\phi_2$ and $\phi_2=\phi_3$, we have to require only that
\be
\label{bc2}
\Psi\Big|_{\phi_2=\phi_1}= {\rm finite}\quad  \wedge  \quad
\Psi\Big|_{\phi_3=\phi_2}={\rm finite}~.
\ee
Finally,
normalisability of the wave functions requires that the wave functions
vanish sufficiently fast at infinity.

\subsection{\normalsize\bf Virial theorem}
Writing $H=\frac{1}{2}\left(E^2+B^2\right)$ and denoting
the eigenstates of H by $|n\rangle $, with energies $E_n$,
one obtains the virial theorem
\be
\label{VirialTh}
\langle n| E^2|n \rangle = 2 \langle n| B^2|n \rangle~,
\ee
It proofs to be a very useful tool to judge
the quality of the approximate eigenstates obtained using the
variational approach. From (\ref{VirialTh}) also follows
($G^2:= 2\left(B^2-E^2\right)$)
\be
\langle
n| B^2|n \rangle = \frac{2}{3}E_n ~,\ \ \langle n| E^2|n
\rangle = \frac{4}{3}E_n ~,\ \ \langle n| G^2|n \rangle =
-\frac{4}{3}E_n ~.
\ee

\section{\large\bf Low energy spin-0 states}

For ${\bf J}^2=\xi^2=0$, the Hamiltonian (\ref{Hpys}) reduces to
\be
\label{H--0}
H_0\!=\!{1\over 2}\!\sum_{\rm
cyclic}^3\!\! \left[-{\partial^2\over\partial \phi_i^2} -{2\over \phi_i^2-
\phi_j^2}\!\left(\phi_i{\partial\over\partial
\phi_i}-\phi_j{\partial\over\partial \phi_j}\right)\!\! +\!  g^2 \phi_j^2 \phi_k^2\right]
\ee
Similar as in the Calogero model \cite{Calogero} one can prove
that the eigenstates of (\ref{H--0})
are completely symmetric in the arguments $(\phi_1,\phi_2,\phi_3)$.
We therefore can have only the two possible forms, the parity even
\be
\label{even}
\Psi^{(+)}(\phi_1,\phi_2,\phi_3)=
\widetilde{\Psi}^{(+)}(\phi_1^2,\phi_2^2,\phi_3^2)~,
\ee
and the parity odd
\be
\label{odd}
\Psi^{(-)}(\phi_1,\phi_2,\phi_3)=
\phi_1\phi_2\phi_3\widetilde{\Psi}^{(-)}(\phi_1^2,\phi_2^2,\phi_3^2)~,
\ee
where the functions $\widetilde{\Psi}^{(\pm )}$ are completely symmetric
in the arguments $(\phi_1^2,\phi_2^2,\phi_3^2)$.
$\Psi^{(+)}$ satisfies the first and
$\Psi^{(-)}$ the second of the boundary conditions (\ref{bc1}).
I shall consider here only the even case (\ref{even}), lower in energy.
The odd case (\ref{odd}) can be treated completely analogously.

\subsection{\normalsize\bf Elementary symmetric polynomials}
\begin{figure}
\centering
\epsfig{figure=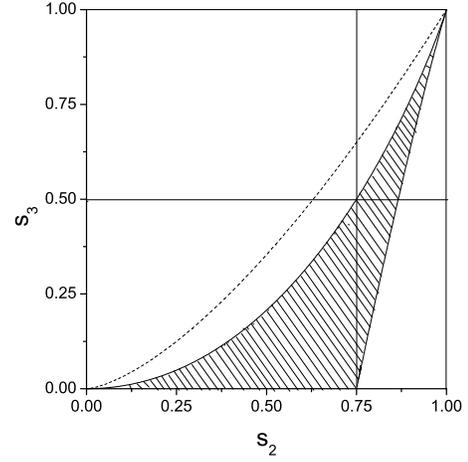,width=60mm}
\caption{\small The (dashed) area of integration in $(s_2,s_3)$ space. The dotted line is the
         limit $s_3\leq s_2^{3/2}$ from algebraic inequalities.}\label{integrarea}
\end{figure}

Here it is useful to  change to the new coordinates $(e_1,e_2,e_3)$
defined as the elementary symmetric combinations
\be
\label{elemsym}
e_1 =\phi_1^2+\phi_2^2+\phi_3^2\ ,\ \ e_2 =\phi_1^2\phi_2^2+\phi_2^2\phi_3^2+\phi_3^2\phi_1^2\ ,
\ \  e_3 =\phi_1^2\phi_2^2\phi_3^2\ .
\ee
The corresponding Jacobian
\be
\label{Jac_e}
\Bigg|{\partial (\phi_1,\phi_2,\phi_3)\over \partial (e_1,e_2,e_3)}\Bigg|
={1\over \sqrt{e_3}\sqrt{\Delta}}~,
\ee
with the square root of the discriminant
\be
\label{discriminant}
\Delta\equiv\! \prod_{i<j}(\phi_i^2-\phi_j^2)^2\!=\!
-27e_3^2\!+\! 18e_1e_2e_3\!-\! 4e_1^3e_3\!-\! 4e_2^3\!+\!e_2^2e_1^2,
\ee
cancels the original measure $\prod_{i<j}(\phi_i^2-\phi_j^2)$ in (\ref{measure}).
Furthermore let us consider the scaling transformation
\be
s_1=e_1~,\ \ \ \ \ \
s_2=3 e_2/e_1^2~,\ \ \ \ \ \
s_3=27 e_3/e_1^3~,
\ee
with the Jacobian $|\partial (e_1,e_2,e_3)/\partial (s_1,s_2,s_3)|\propto s_1^{5}$.
Then the Schr\"odinger equation of (\ref{H--0}) becomes
\be
\label{H--0s}
\left[{1\over 6}\ g^2 s_1^2s_2
-2s_1{\partial^2\over\partial s_1^2}
-9 {\partial\over\partial s_1}
 + {1\over 2 s_1}\Bigg(D^{(0)}
-\frac{49}{4}\Bigg)\!\!\right]\!\Psi=\!E\Psi~,
\ee
with
\be
 D^{(0)}:=\ D_{0}^{(0)}+\ D_{-1}^{(0)}+\ D_{-2}^{(0)}~,
\ee
where
\be
\label{D0}
 D_{0}^{(0)}:=
  \left(2\left(2s_2{\partial\over\partial s_2}
        +3s_3{\partial\over\partial s_3}\right)+{7\over 2}\right)^2~,
\ee
\be
\label{D-1}
 D_{-1}^{(0)}:=
 -4s_3{\partial^2\over\partial s_2^2}
 -18\left(2s_3{\partial\over\partial s_3}
 +1\right)s_2{\partial\over\partial s_3}~,
\ee
\be
\label{D-2}
D_{-2}^{(0)}:= -12\left(s_2{\partial\over\partial s_2}+
4s_3{\partial\over\partial s_3}+2\right)
{\partial\over\partial s_2}~.
\ee
Note that the singularities in $H_0$ have disappeared via the transformation (\ref{elemsym})
to the elementar symmetric variables.
The matrix elements become
\bea
\label{ONRs}
\langle \Psi'|O|\Psi \rangle\!\!\!
&\propto &\!\!\!\!\!\! \int\limits_{0}^{\infty}ds_1\ s_1^{7/2}
\Bigg[
\int\limits_{0}^{3/4} ds_2\!\!\!\!\!\!
\int\limits_{0}^{s_3^{(\rm up)}(s_2)}\!\!\!\!\!\!
{ds_3\over \sqrt{s_3}}
\ \Psi'^{*} O \Psi
\nn\\
&&\quad\quad +
\int\limits_{3/4}^{1} ds_2\!\!\!\!\!\!
\int\limits_{s_3^{(\rm low)}(s_2)}^{s_3^{(\rm up)}(s_2)}\!\!\!\!\!\!
{ds_3\over \sqrt{s_3}}
\ \Psi'^{*} O \Psi
\Bigg]~.
\eea
Here the region  of integration is given by the
$(s_2,s_3)$ satisfying the inequality, obtained from \ref{discriminant},
\be
\tilde{\Delta}\equiv \left(-s_3^2+6s_2s_3
-4s_3-4s_2^3+3s_2^2\right) > 0~.
\ee
Denoting the two roots of the equation $\tilde{\Delta}\left(s_3,s_2\right) = 0$,
quadratic in $s_3$, by
\be
s_3^{(\rm up)}(s_2):=
\left(1-\sqrt{1-s_2}\right)^2
\left(1+2\sqrt{1-s_2}\right)~,
\ee
well defined and positive for all $0\leq s_2 \leq 1/3$, and
\be
s_3^{(\rm low)}(s_2):=
\left(1+\sqrt{1-s_2}\right)^2
\left(1-2\sqrt{1-s_2}\right)~,
\ee
well defined for all $0\leq s_2 \leq 1$, but positive
only for $3/4\leq s_2 \leq 1$.
The region of positive $\tilde{\Delta}$ is the shaded region in Fig. \ref{integrarea}.

\subsection{\normalsize\bf Exact solution of the corresponding harmonic
         oscillator problem}

Note that for the case when the Yang-Mills potential in (\ref{H--0}) is replaced by the
harmonic oscillator potential
\be
g^2 \left(\phi_1^2\phi_2^2+ \phi_2^2\phi_3^2+\phi_3^2\phi_1^2\right)
\longrightarrow
\omega^2 \left(\phi_1^2+ \phi_2^2+\phi_3^2\right)~,
\ee
that is $g^2 s_2s_1^2/6$ replaced by $\omega s_1/2$,
the above Schr\"odinger eigenvalue problem (\ref{H--0s})
separates into a density and a deformation problem\footnote{
Note that the variables $s_1$, $s_2$ and $s_3$ are related to the density
and deformation variables $\rho$,$\beta$ and$\gamma$ used in \cite{Martin}
via $s_1=\rho^2$, $s_2=1-\beta^2$ and $s_3=1-3\beta^2+2\beta^3\cos{\gamma}$.}
($k$ separation const., $\mu$ multiplicity),
\be
\Phi_{n k \mu}(s)=
R_{n k}(s_1)P_{k \mu}(s_2,s_3)~.
\ee
The solutions of the density equations are given by
\be
R_{n k}(s_1)\!=\!
\sqrt{{n!\over\Gamma(n+k+1)}}\ \omega^{9 \over 4}
\left(\omega s_1\right)^{{1\over 2}(k-{7\over 2})}
e^{-\omega s_1/2}L_n^k(\omega s_1)~,
\ee
satisfying the orthonormality relations
\be
\label{ONRdensity}
\int_{0}^{\infty}ds_1\ s_1^{7/2}
\ R_{n k}(s_1)R_{n'k}(s_1)=\delta_{nn'}~,
\ee
with the energy eigenvalues
\be
E_{nk}=\left(2n+1+k\right)\omega~.
\ee
The values of $k$ are determined by the corresponding
deformation problem
\be
\label{deformEqu2}
 D^{(0)} P_{k \mu}(s_2,s_3) =
 k^2\ P_{k \mu}(s_2,s_3)~.
\ee
In the space of monomials $s_2^p\ s_3^q ~,\ \ \ \ p,q\in N_0$ ordered by increasing
$2p+3q$, the operator $D^{(0)}$ has tridiagonal form, since the part
$D_{0}^{(0)}$ is diagonal, and the parts $D_{-1}^{(0)}$ and $D_{-2}^{(0)}$
act as lowering operators, as can be easily seen from (\ref{D0})-(\ref{D-2}).
Hence $D^{(0)}$ can easily be diagonalised with eigenvectors $P_{(p,q)}$
of the form
\be
P_{(p,q)} = \textit{N}_{(p,q)}\left[\ s_2^{p}s_3^{q}
   +\!\!\!\!\!\!\!\!\!\!\sum_{2p'+3q'\leq 2p+3q}
   \!\!\!\!\!\!\!\! a(p, q; p',q')\ s_2^{p'}s_3^{q'}\right],
\ee
and eigenvalues
\be
k=2(2p+3q)+{7 \over 2}~.
\ee
The first $3$ eigenstates $P_{(p,q)}$,
ordered by increasing eigenvalue $k$, are then
\bea
\label{deformstates}
k=\ \ 7/2  &:& P_{(0,0)}= 1~,\\
k=15/2 &:& P_{(1,0)}={11\over 2}\sqrt{13\over 15}
\left(s_2-{6\over 11}\right)~,\nn\\
k=19/2 &:& P_{(0,1)}={221\over 126}\sqrt{209\over 10}\left(s_3-{9\over 17}s_2
     +{36\over 221}\right)~,\nn
\eea
and so on.
Orthonormality of the states $\Phi$ with respect to the scalar product
(\ref{ONRs}), leads to the correponding orthonormality relations for the $P_{k \mu}$,
\bea
&&\!\!\!\!\!\!\!\!
\int_{0}^{3/4} ds_2
\int_{0}^{s_3^{(\rm up)}(s_2)} {ds_3 \over \sqrt{s_3}}
\ P_{k'\mu'} \ P_{k\mu}
\\
&&
+
\int_{3/4}^{1} ds_2
\int_{s_3^{(\rm low)}(s_2)}^{s_3^{(\rm up)}(s_2)}
{ds_3 \over \sqrt{s_3}}
\ P_{k'\mu'}  \ P_{k\mu} ={6\sqrt{3}\over 35}\ \delta_{k'k}\delta_{\mu'\mu}~.\nn
\label{ONRdeform}
\eea
We have choosen the overall normalisation constant
such that the constant solution, $P_{(0,0)}$, corresponding to the lowest eigenvalue $k=7/2$,
is equal to one.
Note that there is no eigenstate with $k=11/2$. To the
eigenvalue $k=31/2$ correspond two states $P_{(3,0)}$ and $P_{(0,2)}$.
One can easily check that states with different values of $k$ are orthonormal
to each other as it should be.
To orthonormalise the states of equal $k$ with respect to (\ref{ONRdeform}),
one can use Gram-Schmidt orthonormalisation, $ P_{(p,q)}\rightarrow P'_{(p,q)}$.
In summary we obtain the eigenstates
\be
\label{ONBho}
\Phi_{n p q}(s)=
R_{n k}(s_1)P'_{(p,q)}(s_2,s_3)~,
\ee
with the energy eigenvalues
\be
E_{n,p,q}=\left(2\left(n+2p+3q\right)+ {9\over 2}\right)\omega~,
\ee
which are equidistant and depend only on the total number $2(n+2p+3q)$ of nodes.
The degeneracy is therefore rapidly increasing with energy.

\subsection{\normalsize\bf Low energy spin-0 spectrum from variational calculation}

To obtain the low energy spectrum of the Hamiltonian
I shall use the Rayleigh-Ritz variational method.
Lead by the parabolic form of the Yang-Mills potential close to the bottom of the
classical zero-energy valley, I shall use the orthonormal basis $\Phi_{n p q}$ in (\ref{ONBho})
as trial functions.
In order to have rapid convergence the frequency is fixed
using the lowest eigenstate
\bea
\Phi_{0 0 0} &=&{\sqrt{\Gamma(9/2)}}\ \omega^{9/4}
                            \exp[-\omega s_1/2]~.
\eea
The stationarity conditions for the energy functional of this state,
\bea
 E[\Phi_{0 0 0}]= \langle \Phi_{0 0 0}|H_0|\Phi_{0 0 0}\rangle \!\!\!\!\!\!&=&\!\!\!\!\!\!
 {9\over 4}\ \omega+
       {9\over 4}\ {g^2\over \omega^2}~,\nn
\eea
leads to the optimal choice
\bea
\label{fr+}
\omega &=& \sqrt[3]{2}\ g^{2/3}\simeq  1.259921 g^{2/3}~.
\label{fr-}
\eea
As a first upper bound for the groundstate energy of the Hamiltonian one
therefore finds
\bea
\label{1stest+}
E_0 \le E[\Phi_{0 0 0}] &\simeq & 4.25223~ g^{2/3}~.
\eea
In order to improve this upper bound,
one truncates the Fock space at a certain number of nodes
of the wave functions and  diagonalizes
the corresponding truncated Hamiltonian $H_{\rm trunk}$ to
find its eigenvalues and eigenstates.
Extending to higher and higher numbers of
nodes I obtain the low energy spectrum
in the spin-0 sector to high numerical accuracy.
Including all $174(1041)$ trial states up to $30(60)$ nodes,
I obtain for the lowest state $S_0$,
\bea
E[S_0] &=& 4.116719740(35)~ g^{2/3}~,
\eea
(i.e. the last two digits have to be replaced by the ones in brackets
when going from $30$ to $60$ nodes).
The state $S_0$, given explicitly as (up to coefficients $< 0.01$)
\bea
S_{0} &=& 0.98\ \Phi_{0 0 0} - 0.07\ \Phi_{1 0 0}
         - 0.04\ \Phi_{2 0 0} \nonumber\\
         &&  -  0.16\ \Phi_{0 1 0}
          + 0.02\ \Phi_{0 0 1} + 0.03\ \Phi_{0 2 0}~,
\label{groundstate}
\eea
nearly coincides with the state $\Psi_{0 0 0}$,
The energies of the groundstate and the first excitations together
with the corresponding deviation from the virial theorem (\ref{VirialTh})
\be
\label{violvir}
\Delta E_{\rm virial} := 2\langle B^2\rangle -\langle E^2\rangle~,
\ee
are shown in Tab. 1.

\begin{table}[b]
\caption{The energies and the deviation from the virial theorem for the lowest
spin-0 eigenstates (using trial states up to 30 (60) nodes).}

\begin{center}
$
\begin{array}{|c|c|c|}
\hline
0^{+}& E [g^{2/3}]  & \Delta E_{\rm virial} [g^{2/3}]\\
\hline
   0  &   4.116719740(35)   &   7.6\cdot 10^{-8}\ (1.2\cdot 10^{-12})  \\
   1  &   6.386361(58)      &   2.4\cdot 10^{-7}\ (1.3\cdot 10^{-10})  \\
   2  &   7.97348(34)       &   4.8\cdot 10^{-4}\ (1.3\cdot 10^{-8})  \\
   3  &   9.2039(23)        &   1.8\cdot 10^{-3}\ (3.1\cdot 10^{-7})  \\
   4  &  10.092(86)         &   7.5\cdot 10^{-4}\ (1.9\cdot 10^{-6})  \\
   5  &  10.966(37)         &  -1.8\cdot 10^{-2}\ (1.8\cdot 10^{-5})  \\
   6  &  12.17(05)          &  -0.22            \ (1.5\cdot 10^{-4})  \\
\hline
\end{array}
$
\end{center}
\end{table}

The energy spectrum of the first spin-0 eigenstates
(with trial states up to 30 nodes) are shown in Fig. \ref{energy}.
The errorbars are inside the lines.
The energies obtained here are in good agreement with those obtained
by L\"uscher and M\"unster \cite{Luescher and Muenster}
and by Koller and van Baal \cite{Koller and van Baal},
using gauge invariant wave functions as trial states.

\section{\large\bf Higher spin states}

For non-vanishing spin we can write the Hamiltonian (\ref{Hpys})
\bea
\label{HHHq}
H=H_0+{1\over 2}\sum_{i=1}^3\xi^2_i V_i~,\ \
V_i={\phi_j^2+\phi_k^2\over (\phi_j^2-\phi_k^2)^2},\ \ {\rm i,j,k\ \ cyclic}\nn
\eea
with the spin-0  Hamiltonian (\ref{H--0}).
The eigenstates can be classified according to their ${\bf J}^2$ and $J_3$ quantum
numbers and are superpositions of simultaneous eigenstates of ${\bf J}^2$,$J_3$ and $\xi_3$
\be
|J M\rangle=\sum_{M'=0,1\pm,..,J\pm} \Psi^{(J)}_{MM'}(\phi_1,\phi_2,\phi_3)|JMM'\rangle~,
\ee
with
the combinations (for $M' > 0$)
\be
|J M M'\!\pm \rangle :=
{1\over \sqrt{2}}\left(|J M M'\rangle \pm|2 M -\!\!M'\rangle \right)~,
\ee
of
\be
|J M M'\rangle := i^J\sqrt{2J+1\over 8\pi^2}D^{(J)}_{M M'}(\alpha,\beta,\gamma)~.
\ee

\subsection{\normalsize\bf Spin 1}

The spin-1 Schr\"odinger equation decays into
three equations, one for each member of the cyclic triplet
$( \Psi_{1}^{(1)},\Psi_{2}^{(1)},\Psi_{3}^{(1)})
= (\Psi_{M 1-}^{(1)},\Psi_{M 1+}^{(1)},\Psi_{M 0}^{(1)})$,
\bea
\left[H_0-E+\frac{1}{2}(V_2+V_3)\right]\Psi_{1}^{(1)}=0~,\quad {\rm and\ cycl.\ perm.}\nn
\eea
One can easily show that no solutions exist which satisfy the boundary conditions (\ref{bc2}).

\subsection{\normalsize\bf Spin 2}

The spin-2 Schr\"odinger equation decays into
and into three equations, one for each member of the cyclic triplet
$( \Psi_{1}^{(2)},\Psi_{2}^{(2)},\Psi_{3}^{(2)})
= (\Psi_{M 1+}^{(2)},\Psi_{M 1-}^{(2)},\Psi_{M 2-}^{(2)})$,
\bea
\left[H_0-E+\frac{1}{2}(V_2+V_3)+2V_1\right]\Psi_{1}^{(2)}=0~,\quad {\rm and\ cycl.\ perm.}\nn
\eea
for which no solutions exist which satisfy the boundary conditions (\ref{bc2}),
and the coupled system
\bea
\label{02+}
\left[\!H_0\!-\!E\!+\!\frac{3}{2}(V_1+V_2)\!\right]\!\Psi_{M 0}^{(2)}
\!+\!\!\frac{\sqrt{3}}{2}(V_1-V_2)\!\Psi_{M 2+}^{(2)}\!\!\!\!\!\!&=&\!\!\!\! 0\nn\\
 \left[\! H_0\!-\!E\!+\!2V_3\!+\!\frac{1}{2}(V_1+V_2)\!\right]\!\Psi_{M 2+}^{(2)}
 \!\!+\!\!\frac{\sqrt{3}}{2}(V_1-V_2)\!\Psi_{M 0}^{(2)}\!\!\!\!\!\!&=&\!\!\!\! 0\nn
\eea
for
$\Psi_{M 0}^{(2)}$ and $\Psi_{M 2+}^{(2)}$.
The solution of this spin-2 singlet system (\ref{02+}) can be written in the form
\bea
\!\!\!\!\!\!\!\!
|2 M\rangle \!\!\!\!&\!\!\!\! =\!\!\!\! &\!\!\!\!\Psi_1(s_1,s_2,s_3)s_1^{-1} Y_M (\phi_1^2,\phi_2^2,\phi_3^2;\alpha,\beta,\gamma)\nn\\ &&+
\Psi_2(s_1,s_2,s_3)s_1^{-2}\widetilde{Y}_M (\phi_1^2,\phi_2^2,\phi_3^2;\alpha,\beta,\gamma)~,
\eea
with the cyclic symmetric functions $\Psi_{1,2}(s_1,s_2,s_3)$ and elementary spin-2 fields
$Y_M (\phi_1,\phi_2,\phi_3;\alpha,\beta,\gamma)$
\be
Y_M\! =\!
\sqrt{2\over 3}\!\left[\!\!\left(\!\phi_3\!-\!\frac{1}{2}(\phi_1\!+\phi_2)\!\!\right)\!\!|2 M 0\rangle
\!+\!\!\frac{\sqrt{3}}{2}\!\left(\phi_1-\phi_2\right)\! |2 M 2+\rangle\!\right]
\ee
and its dual $\widetilde{Y}_M:= Y_M\big|_{\phi_1\rightarrow\phi_2\phi_3} $.
The matrix element of an operator $O(s)$ can be written as
\bea
\langle 2 M'|O| 2 M \rangle\!\!\!\!\!\! &=&\!\!\!\!\!\!\!\!\Big[
\langle \Psi'_1|(\!1\!-\!s_2\!)O|\Psi_1\rangle
+{1 \over 6}\langle \Psi'_1|(s_3-s_2)O|\Psi_2\rangle\nn\\
&&\!\!\!\!\!\!\!\!\!\!\!\!\!\!\!\!\!\!\!\!\!
+{1 \over 6}\langle \Psi'_2|(s_3-s_2)O|\Psi_1\rangle +
{1 \over 9}\langle \Psi'_2|(s_2^2-s_3)O|\Psi_2\rangle\Big]~.\nn
\eea
in terms of the corresponding spin-0 matrix elements (\ref{ONRs}).
Specifying again to parity even states, the vector\newline ${\bf \Psi}=(\Psi_1,\Psi_2)$
satisfies the Schr\"odinger equation
of the same form as (\ref{H--0s}) only with the scalar $D^{(0)}$ replaced with the new matrix operators
$D^{(2)}$ with
\bea
D_{0}^{(2)}\!\!\!\!\!\!&=&\!\!\!\!\!\!\left(\!\!\!\!\begin{array}{cc}
\left(2\left(2s_2{\partial\over\partial s_2}
        +3s_3{\partial\over\partial s_3}+1\right)+{7\over 2}\right)^2
& 0 \\ \!\!\!\!\!\!\!\!\!\!\!\!\!\!\!\!\!\!\!\!\!\!\!\!\!\!\!\!\!\!\!\!\!\!\!\!\!\!\!\! 0
& \!\!\!\!\!\!\!\!\!\!\!\!\!\!\!\!\!\!\!\!\!\!\!\!\!\!\!\!\!\!\!\!\!\!\!\!\!\!\!\!\!\!\!\!\!\!\!\!\!
           \left(2\left(2s_2{\partial\over\partial s_2}
        +3s_3{\partial\over\partial s_3}+2\right)+{7\over 2}\right)^2\!\!\!\!
\end{array}\right),\nn\\
 D_{-1}^{(2)}\!\!\!\!\!\!&=&\!\!\!\!\!\!\left(\begin{array}{cc}
  D_{-1}^{(0)}&
 8s_3{\partial\over\partial s_3}+4\\  24 {\partial\over\partial s_2}
 &  D_{-1}^{(0)}
\end{array}\right)~,\nn\\
 D_{-2}^{(2)}\!\!\!\!\!\!&=&\!\!\!\!\!\!\left(\begin{array}{cc}
  D_{-2}^{(0)} & 0 \\ 0 &
  D_{-2}^{(0)} -24 {\partial\over\partial s_2}
\end{array}\right)~.\nn
\eea
Again the corresponding harmonic oscillator Schr\"odinger Equation separates
\be
{\mathbf \Phi}_{n k \mu}= R_{n k}(s_1){\bf P}_{ k \mu}(s_2,s_3)~.
\ee
The density equations for $R_{nk}$ are the same as in the spin-0 problem
The values of $k$ are determined by the corresponding
deformation problem
\be
 D^{(2)} {\bf P}_{ k \mu}(s_2,s_3)=k^2 {\bf P}_{ k \mu}(s_2,s_3) ~,
\ee
using the basis $(s_2^p s_3^q,0)$ and $(0,s_2^{\tilde{p}} s_3^{\tilde{q}})$
The first $3$ eigenstates $P_{(p,q)}$,
ordered by increasing eigenvalue $k$, are then
\bea
k=11/2 &:& {\bf P}_{(0,0)}= {2\over \sqrt{195}}{1\choose 0}~,\nn\\
k=15/2 &:& {\bf P}_{(\tilde{0},\tilde{0})}=\sqrt{26\over 35}{2/13\choose 1}~,
\nn\\
k=19/2 &:& {\bf P}_{(1,0)}={17\over 7}\sqrt{19\over 15}{s_2-6/17\choose 12/17}
     ~,\nn
\eea
and so on.

As for the spin-0 case one can use this orthonormal basis of spin-2 states of
the harmonic oscillator problem as trial states for a variational calculation
of the corresponding eigenstates of the spin-2 Yang-Mills Hamiltonian.
The energies of the groundstate and the first excitations together
with the corresponding deviation from the virial theorem (\ref{VirialTh})
are shown in Tab. 2.

\begin{table}[b]
\caption{The energies and the deviation from the virial theorem for the lowest
spin-2 eigenstates (using trial states up to 30 nodes).}

\begin{center}
$
\begin{array}{|c|c|c|}
\hline
2^{+}& E [g^{2/3}]  & \Delta E_{\rm virial} [g^{2/3}]\\
\hline
   1  &  6.014500       &   6.3\cdot 10^{-6}   \\
   2  &  7.82062        &   4.8\cdot 10^{-4}   \\
   3  &  9.3335        &   4.2\cdot 10^{-3}   \\
   4  &  9.9285         &   6.0\cdot 10^{-4}   \\
   5  &  10.812         &  -3.0\cdot 10^{-2}   \\
   6  &  11.91          &  -0.10               \\
\hline
\end{array}
$
\end{center}
\end{table}

\begin{figure}
\centering
\epsfig{figure=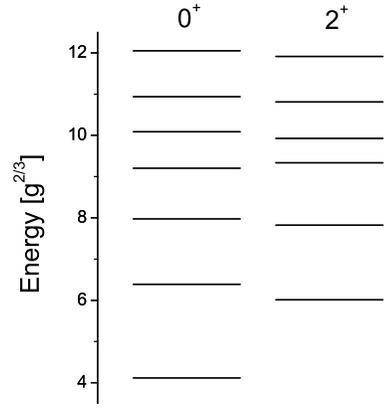,width=50mm}
\caption{\small The energy spectrum of the first spin-0 and spin-2 eigenstates.}\label{energy}
\end{figure}

The energy spectrum of the first spin-2 eigenstates
(including the $270$ trial states with up to $30$ nodes) is shown in Fig. \ref{energy}.
The errorbars are inside the lines.
Also for spin-2 the energies obtained here are in good agreement with those obtained
by L\"uscher and M\"unster \cite{Luescher and Muenster}
and by Koller and van Baal \cite{Koller and van Baal}.
In particular I confirm that the first spin-2 state is lower in energy than the first
spin-0 excitation.

\section{\large\bf Expectation values of permutation non-invariant operators}

\begin{figure}[t]
\centering
\epsfig{figure=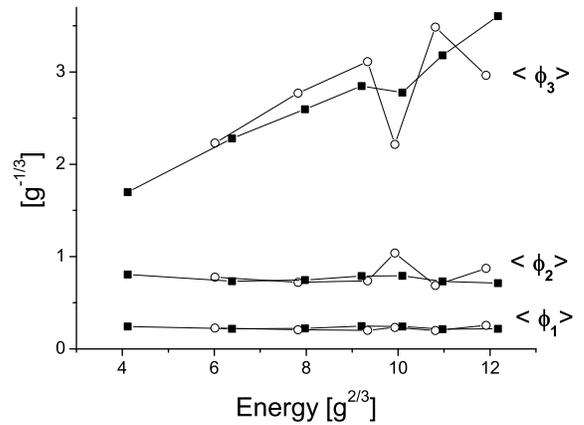,width=75mm}
\caption{\small The expectation values $\langle \phi_1\rangle$, $\langle \phi_2\rangle$
         and $\langle \phi_3\rangle$ of the scalar fields along the principal axes
         as function of the energy for the lowest spin-0 (black boxes) and
         spin-2 (open circles) states.}\label{Fig. 3}
\end{figure}

Expanding the obtained energy eigenstates in terms of the original $\phi_1,\phi_2,\phi_3$
and analytically calculating the Gaussian integrals in the region (\ref{range}),
we obtain the expectation values $\langle \phi_1\rangle$,$\langle \phi_2\rangle$ and
$\langle \phi_3\rangle$ of the three fields along the principal axes.
For the spin-0 grounstate e.g. we obtain
\bea
\langle 0|\phi_1|0\rangle_+ &=& 0.242~g^{-1/3},\nn\\
\langle 0|\phi_2|0\rangle_+ &=& 0.806~g^{-1/3},\nn\\
\langle 0|\phi_3|0\rangle_+ &=& 1.699~g^{-1/3}.\nn
\eea
Fig. \ref{Fig. 3} shows the results
for the lowest spin-0 (dark symbols) and spin-2 states (open white symbols)
as function of the energy of the states.
I find that $\langle \phi_3\rangle$ raises universally with increasing energy,
independent whether oscillator or spin excitation,
whereas $\langle \phi_1\rangle$ and $\langle \phi_2\rangle$ stay practically
constant.

\begin{figure}[t]
\centering
\epsfig{figure=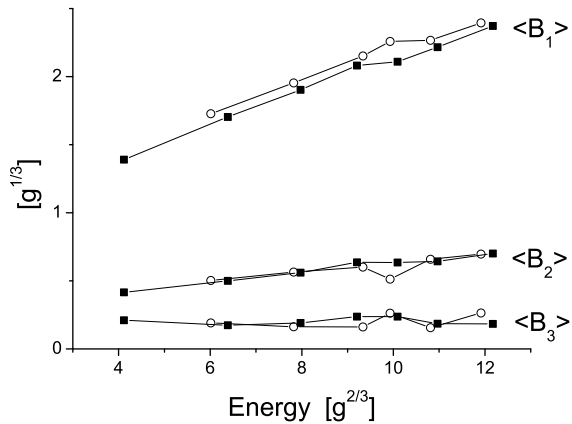,width=75mm}
\caption{\small  The expectation values $\langle B_1\rangle$, $\langle B_2\rangle$
         and $\langle B_3\rangle$ of the magnetic fields along the principal axes
         as function of the energy for the lowest spin-0 (black boxes) and
         spin-2 (open circles) states.}\label{Fig. 4}
\end{figure}

The corresponding expectation values  of the three
intrinsic magnetic fields $B_1=g\phi_2\phi_3$ etc.
\bea
\langle 0|B_1|0\rangle_+ &=& 1.391~g^{1/3}~,\nn\\
\langle 0|B_2|0\rangle_+ &=& 0.415~g^{1/3}~,\nn\\
\langle 0|B_3|0\rangle_+ &=& 0.211~g^{1/3}~,\nn
\eea
in accordance with the opposite ordering $B_1>B_2>B_3$, following from (\ref{range}).
The results for $\langle B_1\rangle$,$\langle B_2\rangle$ and $\langle B_3\rangle$
as function of the energy for the lowest spin-0 and spin-2 states are shown in Fig. \ref{Fig. 4}.
$\langle B_3\rangle$ stays practically constant with increasing energy,
whereas $\langle B_1\rangle$ and $\langle B_2\rangle$ are raising.

Up to the considered number of $30$ nodes the numerical errors are much smaller
in both cases than the systematical error of the obtained states ($ \leq 1 \%$).

\section{\large\bf Summary and discussion}

It has been shown that $SU(2)$ Yang-Mills quantum mechanics
can be studied very effectively and with high accuracy, by choosing the symmetric gauge
and performing a variational calculation using the orthonormal basis of exact eigenstates
the corresponding harmonic oscillator problem as trial states
in the spin-0 and the spin-2 sector.
The virial theorem is shown to be fulfilled very well for the eigenstates obtained.
The energies obtained here are in good agreement with those obtained
by L\"uscher and M\"unster \cite{Luescher and Muenster}
and by Koller and van Baal \cite{Koller and van Baal},
using gauge invariant wave functions as trial states.
In particular I confirm that the first spin-2 state is lower in energy than the first
spin-0 excitation.
Furthermore I have found that practically all excitation energy, independently whether it is due to
a vibrational or a rotational excitation, goes into the increase of the field strength along
one of the three principal axis fields, $\langle\phi_3\rangle$,
whereas $\langle\phi_1\rangle$ and $\langle\phi_2\rangle$, and
also the component $\langle B_3\rangle$ of the magnetic field along the intrinsic $3$-direction,
remain unchanged at their vacuum values.

Several techniques in my treatment,
such as the separation of the harmonic oscillator problem into
a density and a deformation problem
and the diagonalization of the deformation operator
in terms of symmetric polynomials,
have for the case of spin-0 to some extent already been developed
in the context of the Calogero model \cite{Calogero}
and then later also in the context of Yang-Mills quantum mechanics in \cite{Koller and van Baal}.
In difference to \cite{Koller and van Baal}, the range of integration (\ref{range}) here
is positive definite, and the corresponding range and measure (\ref{ONRs}) in terms of
the elementary symmetric variables is derived explicitly from the original measure (\ref{measure}).
In particular the relative normalization of the deformation
states (\ref{deformstates}) is different to that given in \cite{Koller and van Baal}.
Furthermore the generalization of the use of
the elementary symmetric variables to spin-2 and the consideration of permutation non-invariant
expectation values here is new to the best of my knowledge.

\section*{\large\bf Acknowledgments}

I would like to thank A. Dorokhov, J. Wambach, and  W. Weise for their interest and support.
Financial support by the GSI-Darmstadt is gratefully acknowledged.


\end{document}